# PhyloGrid: a development for a workflow in Phylogeny

Esther Montes[1], Raúl Isea[2], Rafael Mayo[1]

[1] CIEMAT, Avda. Complutense, 22, 28040 Madrid, Spain
[2] Fundación IDEA, Centro de Biociencias, Hoyo de la Puerta, Valle de Sartenejal, Baruta 1080, Venezuela

Abstract

In this work we present the development of a workflow based on Taverna which is going to be implemented for calculations in Phylogeny by means of the MrBayes tool. It has a friendly interface developed with the Gridsphere framework. The user is able to define the parameters for doing the Bayesian calculation, determine the model of evolution, check the accuracy of the results in the intermediate stages as well as do a multiple alignment of the sequences previously to the final result. To do this, no knowledge from his/her side about the computational procedure is required.

Keywords.

MrByes, Phylogeny, Taverna, Workflow

Introduction

The determination of the evolution history of different species is nowadays one of the more exciting challenges that are currently emerging in the computational Biology [1,2], where Phylogeny is able to determine the relationship among the species and, in this way, to understand how the influence of the host is affecting it. As an example, we can mention the work from Korber et al. [3] where the evolution history of the AIDS disease was determined. Among others, it was also deduced that AIDS doesn't become from a contaminated sample of a polio vaccine in Africa [4], but it had been brought up since many years ago.

In biological research, scientists often need to use the information of the species which infers the evolutionary relationship among them by means of representations of binary trees, called the evolutionary or phylogenetic trees. Reconstructing evolutionary tree is a major research problem in Biology, the difficulty of which lays on the fact that the number of possible evolutionary trees is very large. As the number of species increases, exhaustive enumeration of all possible relationships is not feasible. The quantitative nature of species relationships therefore requires the development of more rigorous methods for tree construction. As an example, 30 sequences can generate around $8.69 \cdot 10^{36}$ trees with different topology [Krane02]

Several techniques for estimating phylogenetic trees have been developed such as distance based, maximum parsimony, maximum likelihood and bayesian methods (mainly Markov Chain Monte Carlo inference). In this work, the latest will only be used with help of MrBayes software [6] for obtaining the phylogenetic trees. It is important to indicate that MrBayes is relatively new in the construction of these trees as the reader can check in the pioneering work of Rannala and Yang in 1996 [7]. This methodology works with the Bayesian statistics previously proposed by Felsentein in 1968 as indicated Huelsenbeck [8], a technique for maximizing the subsequent probability. The reason for using this kind of approach is that it deals with higher computational speed methods so the possible values for the generated trees can all be taken into account not being any of them ruling the others.

The phylogeny problem is then computationally intensive, thus it is suitable for distributed computing environment. Grid computing is already integrating the CPU power, the storage and other resources via Internet in order to get overall computing power and reduce the execution time. Designs and developments of grid-based systems, which propose efficient methods for solving the Phylogeny problem on this architecture, can be found in the literature [Liu04, Keane05, Minh07].

With the advent of Grid and application technologies, scientists and engineers are building more and more complex applications to manage and process large data sets, and execute scientific experiments on distributed resources. Such application scenarios require means for composing and executing complex workflows (we can define them as a set of components and relations between them used to define a complex process from simple building blocks). Therefore, many efforts have been made towards the development of workflow management systems for Grid computing [Yu05, Oinn05, Chua05] since it is a challenge to integrate new computational techniques that have been previously developed with a different frame of mind.

Because of this huge amount of new services and possibilities for the final researchers that could drive them to reject such a complexity, a more friendly environment had to be developed in order to attract the scientific community to the use of these new advances in a simple and unique interface. Thus, the web portals overcome the problem making easy to the users the calculations on Grids avoiding them to use the command line. The Bioinformatics field was one of the first to join this philosophy and create a wide range of web portals. Some publications related to this topic are [Budura07, Li07], but a wide range of them can be found in Internet.

As a consequence, the aim of this work is to join both aspects of Science (Phylogeny) and Technology (Grid workflows in a web portal) creating a new application that offers to the final user a friendly and easy way to obtain accurate biological results. It is done by means of a simple interface and allowing the user to validate the accuracy of the results during the intermediate stages. This latest point is extremely important because it prevents the production of automatic results that could be flimsy from the biological point of view if the job execution has not been supervised by an expert.

The tools

*MrBayes*

Bayesian inference is a powerful method which is implemented in the program MrBayes [Ronquist03] for estimating phylogenetic trees that are based on the posterior probability distribution of the trees. Comparing to the parsimony and distance methods, Bayesian inference takes full advantage of the information contained in the alignment of DNA sequences (even can make use of morphological data) when estimating phylogenies. Due to the nature of Bayesian inference, the simulation can be prone to entrapment in local optima. To overcome local optima and achieve better estimation, the MrBayes program has to run for millions of iterations (generations) which require a large amount of computation time. For example, the phylogenetic estimation for a medium size dataset (50 sequences, 300 nucleotides for each sequence) typically requires a simulation for 250,000 generations, which normally runs for 50 hours on a PC with an Intel Pentium4 2.8 GHz processor. For multiple sessions with different models or parameters, it will take a very long time before the results can be analyzed and summarized.

There is a MPI-enabled parallel version of MrBayes [Altekar04] that can be compiled and installed in any Linux cluster. Performance results show there is nearly a linear speed up for the parallel jobs. The same job which originally took two days can be finished within few hours. The significant improvement of simulation speed also allows to run the application for a longer session to achieve more stable and accurate estimation. Moreover, the job submission through LSF queues is very flexible and convenient for managing jobs.

*Gridsphere*

The primary goals of the Gridsphere project [gridsphere], developed by a consortium mainly formed by the the University of California-San Diego, the Poznan Supercomputing and Networking Center and the Albert Einstein Institute but with many other collaborators from other Institutions, are to develop a standards based portlet framework for building web portals and a set of portlet web applications that work seamlessly with the GridSphere framework to provide a complete Grid portal development solution. It is based on Grid computing and integrates into the GridSphere portal framework a collection of gridportlets provided as an add-on module, which forms a cohesive "grid portal" end-user environment for managing users and provides access to information services. Grid security based on public key infrastructure (PKI) and emerging IETF and OASIS standards are also well-defined characteristics. The Grid Portlets web application enables users to upload their Grid credentials and use them to gain access to a variety of Grid services. GridSphere provides an implementation of the JSR 168 portlet API standard and a key feature of the design of GridSphere is that it builds upon the web application repository (WAR) deployment model to support third-party portlet web applications.

*Taverna*

Taverna Project [taverna] is a collaboration led by the European Bioinformatics Institute that allows users to construct complex analysis workflows from components located on both remote and local machines, run these workflows on their own data and visualise the results. To support this core functionality it also allows various operations on the components themselves such as discovery and description and the selection of personalised libraries of components previously discovered to be useful to a particular application. Taverna is based on workbench windows where a main section of an Advanced Model Explorer shows a tabular view of the entities within the workflow. This includes all processors, workflow inputs and outputs, data connections and coordination links. For using Taverna, it is necessary to create a web service for the required application that will be integrated into the software. Lately, this web service will call MrBayes inside the workflow.

The workflow

The workflow is fully built in Taverna and structured in different services that are equivalent to the different sections that are run in a common MrBayes job and performs a complete calculation just building the input file by means of the construction of a common Grid `jdl` file. The front end to the final user is a web portal built upon the Gridsphere framework. This solution makes very easy to the researcher the calculations of molecular phylogenies by no typing at all any kind of command.

The main profit that these kind of workflows offer is that they integrate in a common development several modules (tools) connected one to each other. They also allow the user to validate the calculations as the phylogenetic tree is being constructed with the help of visual check-outs, so a higher reliability on the final results is expected. Even more, the deployment of the application with Taverna allows any researcher to download the complete workflow, making easy in this way their spread in the scientific community.

Another advantage of this work is the fact that it groups in a common interface the different tools that are available for putting in the required format the input sequences, i.e. NEXUS for MrBayes

The schema of PhyloGrid is explained in the following paragraphs.

It starts with a common log-in page, which conducts the researcher to his/her user page where the previous calculations together with their status (in progress/complete) and their descriptions previously written by the user are available as well as a "Help" functionality. The way that this process is provided is by means of the personal Grid user certificate ("myproxy" initialization), the execution of which is already integrated in the Gridsphere release. There is also the possibility of running the jobs with a proxy provided by the Administrator that would be renewed from time to time in order to allow longer jobs to be finished.

When the user wants to start a new submission, then a new window is open. In this page, he/she is able to define the name of the job to be submitted as well as its description, to upload the file with the alignment, to select the model of evolution and the number of iterations with a fixed frequency and, finally, to run the experiment. All of this is the usual way of running MrBayes with a basic master file that can be written this way:

```
begin mrbayes;
set autoclose=yes nowarn=yes;
execute primates.nex;
lset nst=6 rates=gamma;
mcmc nruns=1 ngen=10000 samplefreq=100 file=primates.nex1;
mcmc file=primates.nex2;
mcmc file=primates.nex3;
end;
```

so the workflow must perform its load section by section. Since the first two instructions are always the same for any kind of calculation, the workflow has to begin with the third one (`execute`...) making a call for aligning the sequences to be studied. In this point and for a future release, the workflow must be able to translate to a NEXUS format the input alignment if it is written in any other kind of format (clustal, phylips, MSA...).

A definition of the kind of Phylogeny to be used copying the MrBayes model is made in the fourth instruction (`lset`...). Here, two options are possible: to allow the researcher to select a specific one or to allow the workflow to do so. In a first step, we have chosen the first option leaving the second one for a future release with the opensource program ReqdSeq [readseq].

The following instruction (`mcmc nruns`...) sets the number of executions to be made –10000 in our example– with a concrete frequency –100– and the name of the output files that are going to be generated, which will be able to be monitored thanks to the `mcmc` command. All these options are able to be changed by the final user, who at the beginning of the process has simply defined the name of the output files since MrBayes sets the corresponding extensions for them (`.p`, `.t` and `.mcmc`).

A point that has been dropped out in this previous text is the alignment of the sequences. If to the user's knowledge they already are, the workflow moves to a next window where the user is able to choose between writing the output in a MrBayes format or not and, after that, the job is submitted for the Bayesian calculation. If the sequences are not (sufficiently) aligned, they can be re-aligned submitting a MPI-Clustal job, which is monitored in the same webpage and is able to show the result by means of the Jalview tool [jalview].

This calculation can be redefined and iterated by the user at his/her convenience and, in Biological terms, it is very important because the accuracy of the phylogenetic tree, which could be obtained with MrBayes or any other tool, depends on the accuracy of the previously performed alignment. That is why the inclusion of the Jalview tool it is very important from the biological point of view.

This program shows the quality of the alignment (by means of a simple applet) and allows the user to change its structure.

If we also take into account that the spirit of PhyloGrid is to work with a high number of sequences, it is important to count on the availability of tools which reduce as much as possible the computational time, so the re-alignment of those sequences with MPI-Clustal is an asset.

Once the workflow has started, MrBayes automatically validates the number of iterations meanwhile it begins to write the output file and sets the burning value for generating the phylogenetic trees. Once the whole calculation is ended, it can be downloaded by the user for further analysis.

Conclusion and future work

In this work, we present the development of a simple workflow to calculate phylogenetic trees by means of MrBayes tool. It is also integrated in a web portal making easy in this way its use by a any kind of researcher.

We have aforementioned that new functionalities are planned to be incorporated to the workflow in a future release, i.e. translations to NEXUS format for any other usual input files or automatic generation of the model of the Phylogeny.

Nevertheless, a more ambitious future step is to add new phylogenetic tools, such as PBPI, offering to the user a wide range of possibilities to perform different phylogenies. This is important since the capabilities of the programs are not always overlapped, so, for example, depending on the required accuracy of the results MrBayes or PBPI [Feng06] are more convenient. Another clear case for the use of one or the other tool is the number of sequences and their length: the higher they are, the more convenient that PBPI is.

This future integration will deal with the fact that PBPI is based on XML establishing a great difference with MrBayes since it is based on simple text.

Of course that further work will be of interest: analysis routines via web, LDAP management, other portal frameworks such as P-GRADE [pgrade] if necessary, etc. but they are up to now clearly out of the scope of this work.

Acknowledgments

Authors thank in particular the support provided by the EELA-2 Project (E-science grid facility for Europe and Latin America, http://www.eu-eela.org), Grant Agreement n° 223797 of the EU Seventh Research Framework Programme-Research Infrastructures.